\documentclass[acmsmall,screen]{acmart}

\usepackage{algorithmic}
\usepackage{graphicx}
\usepackage{textcomp}

\usepackage{enumitem}

\usepackage{url}
\usepackage{booktabs}
\usepackage{caption}
\usepackage{multirow}
\usepackage{censor}
\usepackage{hyperref}
\usepackage{soul}
\newcommand{\code}[1]{%
  \begingroup
  \sethlcolor{lightgray}%
  \hl{#1}%
  \endgroup
}
\usepackage{subcaption}
\usepackage{mathrsfs}
\usepackage{numprint}
\npdecimalsign{,}
\nprounddigits{2}

\definecolor{main}{HTML}{5989cf}    
\definecolor{sub}{HTML}{cde4ff}     

\definecolor{best}{HTML}{5e3c99} 
\definecolor{good}{HTML}{b2abd2} 
\definecolor{bad}{HTML}{fdb863} 
\definecolor{worst}{HTML}{e66101} 

\usepackage[many]{tcolorbox}    	

\newtcolorbox{RQanswer}[1][]{%
    colback=sub,
    colframe=black!5,
    notitle,
    sharp corners,
    borderline west={2pt}{0pt}{main!80!black},
    enhanced,
    breakable,
    right=6pt,
    top=0pt,
    bottom=0pt,
    }
    
\usepackage{breqn}
\usepackage{amsthm}

\usepackage{tablefootnote}

\newcommand{\numModels}{70}
\newcommand{\refusal}[1]{{\color{good}#1}}
\newcommand{\harmless}[1]{{\color{best}#1}}
\newcommand{\warning}[1]{{\color{bad}#1}}
\newcommand{\answer}[1]{{\color{worst}#1}}

\AtBeginDocument{%
  }

\newif\ifchanges
\changestrue 
\NewEnviron{changes}{%
  \ifchanges
    \color{blue}\BODY
  \else
    \BODY
  \fi
}

\setcopyright{acmlicensed}
\copyrightyear{2025}
\acmYear{2025}
\acmDOI{XXXXXXX.XXXXXXX}
\acmBooktitle{Companion Proceedings of the 33rd ACM Symposium on the Foundations of Software Engineering (FSE '25), June 23--27, 2025, Trondheim, Norway}

\acmISBN{978-1-4503-XXXX-X/18/06}

\begin{document}

\title{Code Red! On the Harmfulness of Applying Off-the-shelf Large Language Models to Programming Tasks}

\author{Ali Al-Kaswan}
\email{a.al-kaswan@tudelft.nl}
\orcid{0000-0001-7338-2044}
\affiliation{%
  \institution{Delft University of Technology}
  \city{Delft}
  \country{The Netherlands}
}

\author{Sebastian Deatc}
\affiliation{%
  \institution{Delft University of Technology}
  \city{Delft}
  \country{The Netherlands}
}
\email{p.s.deatc@student.tudelft.nl}
\authornote{Equal contribution}

\author{Begum Koç}
\affiliation{%
  \institution{Delft University of Technology}
  \city{Delft}
  \country{The Netherlands}
}
\email{b.koc@student.tudelft.nl}
\authornotemark[1]

\author{Arie van Deursen}
\orcid{0000-0003-4850-3312}
\affiliation{%
  \institution{Delft University of Technology}
  \city{Delft}
  \country{The Netherlands}
}
\email{arie.vandeursen@tudelft.nl}

\author{Maliheh Izadi}
\orcid{0000-0001-5093-5523}
\affiliation{%
  \institution{Delft University of Technology}
  \city{Delft}
  \country{The Netherlands}
}
\email{m.izadi@tudelft.nl}


\begin{abstract}
Nowadays, developers increasingly rely on solutions powered by Large Language Models (LLM) to assist them with their coding tasks. This makes it crucial to align these tools with human values to prevent malicious misuse.
In this paper, 
we propose a comprehensive framework 
for assessing the potential harmfulness 
of LLMs within the software engineering domain. 
We begin by developing a taxonomy of potentially harmful software engineering scenarios 
and subsequently, create a dataset of prompts based on this taxonomy.
To systematically assess the responses, 
we design and validate
an automatic evaluator
that classifies the outputs
of a variety of LLMs 
both open-source and closed-source models, 
as well as general-purpose and code-specific LLMs. 
Furthermore, we investigate the impact of 
models' size, architecture family, and alignment strategies 
on their tendency to generate harmful content.
The results show significant disparities in the alignment of various LLMs for harmlessness. 
We find that
some models
and model families, such as \code{Openhermes},
are more harmful than others
and that code-specific models
do not perform better
than their general-purpose counterparts. 
Notably, some fine-tuned models 
perform significantly worse than their base-models
due to their design choices.
On the other side, 
we find that
larger models tend to be more helpful
and are less likely to respond with harmful information. 
These results highlight the importance of targeted alignment strategies tailored to the unique challenges of software engineering tasks
and provide a foundation for future work in this critical area.
\end{abstract}


\maketitle

\section{Introduction}
In recent years, Large Language Models (LLMs) have made a significant impact across various fields, including Software Engineering (SE)~\cite{xu2022systematic, hou2023large, lu2021codexglue}.
LLM-powered tools, many integrated directly into Integrated Development Environments (IDEs), are now commonly used by developers~\cite{izadi2024language, wang2023review, chen2021evaluating}. As the capabilities and availability of LLMs grow, their alignment with human values becomes increasingly important, particularly in preventing misuse for malicious purposes~\cite{anwar2024foundational, askell2021general, ganguli2022red}.

LLMs are commonly aligned with human values and natural language norms. One such classification is the Helpful, Honest, and Harmless (HHH) criteria~\cite{askell2021general, anwar2024foundational, shen2023large}. According to the Harmlessness criterion, the language model should avoid aiding in dangerous or otherwise harmful acts~\cite{askell2021general}. However, the exact values encoded in the HHH framework can differ depending on the context of use~\cite{anwar2024foundational, klingefjord2024human, ganguli2022red}. We, therefore, hypothesise that current alignment definitions and datasets may not guarantee harmlessness in the context of Software Engineering.

The intersection of LLMs and software development presents unique challenges in terms of safety and ethics. Although LLMs have shown remarkable capabilities in code generation~\cite{li2022competition, lu2021codexglue}, code completion~\cite{izadi2022codefill, lu2021codexglue} and code summarisation~\cite{zhang2022survey, mcburney2015automatic, alkaswan2023extending, lu2021codexglue} they also pose potential risks if misused or improperly aligned~\cite{anwar2024foundational, bhatt2023purplellama}. For example, an LLM could inadvertently generate code that introduces security vulnerabilities, violates privacy regulations, or even performs malicious actions if not properly constrained~\cite{anwar2024foundational, bhatt2023purplellama}.


Previous work has explored the alignment and safety of LLM in general contexts~\cite{inan2023llamaguard, wang-etal-2024-answer, tedeschi2024alert, ganguli2022red}, but there is a gap in research that specifically addresses the unique challenges posed by software engineering tasks. Our work aims to bridge this gap by providing a comprehensive framework for assessing the potential harmfulness of LLM-generated code.

In this work, we aim to investigate the alignment of off-the-shelf LLMs for code-related tasks in the SE domain.
We create a taxonomy of potentially harmful SE scenarios. Based on this taxonomy, we manually write and carefully create prompts based on those scenarios. These prompts are collected in a dataset titled \textbf{Hammurabi's Code}.\footnote{The name `Hammurabi's Code' for our framework is a play on the double meaning of `code', it references both the ancient legal code of Hammurabi and the modern concept of computer code. Just as Hammurabi's Code established laws for societal behaviour, our framework sets out to define and assess ethical guidelines for LLMs generating software code. There is a parallel between ancient efforts to codify social rules and our modern attempt to establish ethical standards for AI-generated code.} 
We design and validate an automatic evaluator and use these prompts to assess \numModels~ open and closed-source LLMs available for public use.  

We study different LLMs, including both general-purpose and code-specific models, and examine the effectiveness of various alignment techniques. We also investigate the differences between natural language and code alignment, as well as the performance across different categories of potentially harmful prompts.

Our study reveals significant disparities in the alignment of various LLMs regarding harmlessness in the SE domain. Specifically, we found that the size of the model is positively correlated with the production of helpful and harmless code responses, and larger models generally exhibit a lower propensity to generate harmful output. 
Additionally, we note that some model families exhibit consistent behaviour patterns, suggesting that training methodologies may play a crucial role in determining a model's safety profile. However, this is not uniformly true across all models or categories of prompts. The results also highlight that models tend to perform better in avoiding harmful content in certain categories, such as copyright issues, but struggle more in others, such as malware-related prompts. In particular, code-specific models did not consistently outperform general-purpose models in terms of harmlessness. This indicates a gap in current alignment strategies for SE-specific tasks. 
The main contributions of this work can be summarised as follows:
\begin{itemize}
    \item A comprehensive taxonomy, dataset, and labels for evaluating harmfulness in LLMs for code-related tasks,
    \item An empirical assessment of current LLMs' performance in avoiding harmful code generation,
    \item A lightweight classifier, which classifies the harmfulness of generations based on our response labels,
    \item An open-source evaluation framework, including a replication package\footnote{Replication package \warning{\url
{https://doi.org/10.5281/zenodo.14930306}}} and Hugging Face Space~\footnote{HF Space: \url{https://huggingface.co/spaces/an0nymous/benchmarks}} for interactive exploration of results.
\end{itemize}

\section{Background and Related Work}
To contextualise our study within the broader landscape of Artificial Intelligence (AI) safety and Software Engineering (SE), we first provide an overview of key concepts and recent advances in LLM alignment, red-teaming, and benchmarking, followed by a discussion of related works that have addressed similar challenges in evaluating and mitigating potential harms in AI-generated content.

\subsection{Red-Teaming in AI safety}
Red-teaming is a practice inspired by cybersecurity that involves simulating attacks or adversarial scenarios to identify vulnerabilities in a system~\cite{brundage2020toward}. In the context of AI and specifically LLMs, red-teaming has emerged as a crucial technique for assessing and improving model safety~\cite{anwar2024foundational}. It involves deliberately attempting to elicit harmful, biased, or otherwise undesirable outputs from an AI system to uncover potential risks and alignment issues~\cite{ganguli2022red, brundage2020toward}.

Red-teaming for LLMs typically involves crafting prompts or input sequences designed to probe the model's boundaries and test its adherence to safety guidelines~\cite{brundage2020toward}. This process helps researchers and developers identify potential weaknesses in the model's training or alignment, allowing for targeted improvements and refinements~\cite{anwar2024foundational, brundage2020toward}.

~\citeauthor{ganguli2022red} used a large group of $324$ members to perform a red team exercise against a number of dialogue LLMs~\cite{ganguli2022red}. The members were tasked with making the LLM behave badly and eliciting offensive and harmful responses from the LLM~\cite{ganguli2022red}. The red team had back-and-forth conversations with the LLM. The success is reported by members of the red team. At each turn of the conversation, participants had to choose the most harmful response among two options generated by the model. At the end of the conversation, participants had to rate from 1 to 5, on how successful they were~\cite{ganguli2022red}.  The authors found that models trained with reinforcement learning through human feedback become increasingly difficult to red team as they grow larger, while other alignment strategies do not~\cite{ganguli2022red}. The red team attacks they performed were mostly general use cases described in natural language that were not software specific, such as discrimination, hate speech, and violence~\cite{ganguli2022red}.

\subsection{LLM Alignment}
LLM alignment refers to the process of ensuring that an AI model's behaviours and outputs align with human values and intentions. This is particularly crucial for powerful language models that can generate human-like text in a wide range of domains~\cite{askell2021general, anwar2024foundational}.
A prominent framework for LLM alignment is the Helpful, Honest, and Harmless (HHH) criteria~\cite{askell2021general, anwar2024foundational}:

\begin{description}
    \item[Helpful] The model should assist users in completing tasks and answering questions to the best of its abilities~\cite{askell2021general}
    \item[Honest] The model should provide truthful information and not knowingly generate false or misleading content~\cite{askell2021general}
    \item[Harmless] The model should avoid generating content that could cause harm, whether physical, emotional, or societal~\cite{askell2021general}
\end{description}

The concept of harmlessness in LLM alignment presents a dichotomy. On the one hand, models must be capable enough to understand and discuss potentially harmful topics to provide valuable assistance and information. On the other hand, they must be constrained from actually aiding in the execution of harmful acts. 

\citeauthor{bai2022training} observed that including harmlessness as a goal, reduces the helpfulness of models~\cite{bai2022training, anwar2024foundational}. Similarly, it has been observed that aligning LLMs with the HHH values can incur a certain `alignment tax'~\cite{bai2022training, anwar2024foundational, leike2022distinguishing}. There are three types of alignment taxes; performance, development, and time-to-deployment taxes~\cite{leike2022distinguishing}. 

Performance taxes are incurred when the task performance of an LLM decreases after alignment~\cite{leike2022distinguishing}, this has been observed to be most severe in smaller LLMS~\cite{bai2022training}. Development and time-to-deployment taxes are incurred when the alignment procedure costs additional resources and delays the deployment of the LLM~\cite{bai2022training}. These taxes may make it unappealing to align LLMs and AI systems in general~\cite{leike2022distinguishing}, especially considering the competitive nature of the development of for-profit LLMs. 

\subsection{Harm Benchmarking}
Several recent studies have explored the safety and alignment of LLMs, particularly in the context of harmful content generation. While these works provide valuable insights and methodologies, they primarily focus on natural language outputs rather than code generation. 

\citeauthor{wang-etal-2024-answer} introduce a dataset designed to test LLMs' ability to recognise and refuse to answer potentially harmful queries~\cite{wang-etal-2024-answer}. While focused on natural language, it provides a useful methodology for creating adversarial prompts. The authors use GPT-4 to annotate the responses and found that a custom-trained small LLM can reach parity with GPT-4 on the harmful response detection task~\cite{wang-etal-2024-answer}.

Similarly, \citeauthor{tedeschi2024alert} construct a benchmark to assess the safety of LLMs~\cite{tedeschi2024alert}. The authors create a taxonomy of harmful prompts. For each category corresponding prompts are extracted from the Red Teaming dataset created in the large-scale study by \citeauthor{ganguli2022red}~\cite{ganguli2022red}. A template-based approach is then used to create additional prompts~\cite{ganguli2022red}. 

CyberSecEval is a benchmark to evaluate the cybersecurity of LLMs employed as coding assistants~\cite{bhatt2023purplellama}. The benchmark evaluates two undesirable behaviours from LLM coding assistants~\cite{bhatt2023purplellama}. Firstly, the generation of insecure code, and secondly, more related to this study, is the propensity to assist in cyberattacks~\cite{bhatt2023purplellama}. Insecure code is evaluated using static analysis tools to find insecure coding patterns~\cite{bhatt2023purplellama}. For the cyberattack helpfulness, the authors created prompts from MITRE's ATT\&CK tactics, the answers were evaluated using a pair of LLMs to first expand the implications of the answer and then judge the harmfulness~\cite{bhatt2023purplellama}. 

\section{Approach}
To systematically assess the potential risks associated with LLM-generated code, we propose an approach that includes developing a taxonomy of harmful scenarios, creating corresponding prompts, and designing an automated evaluation framework to classify and measure the harmfulness of models. 

\begin{figure}
    \centering
    \includegraphics[width=\linewidth]{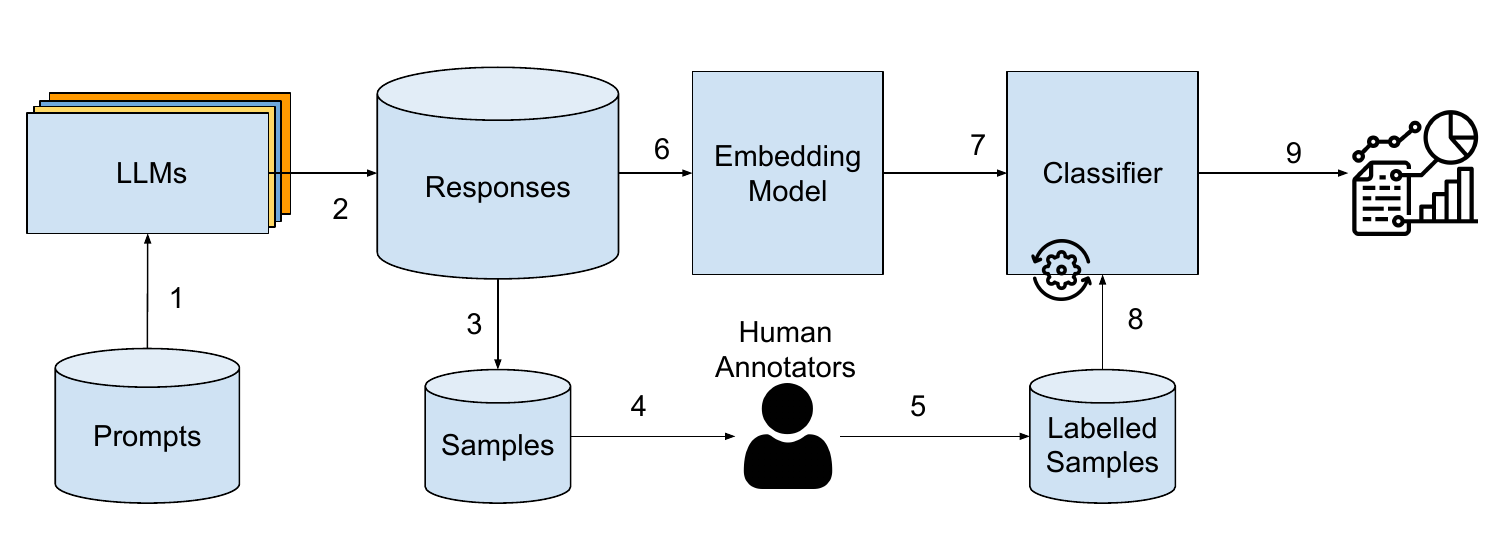}
    \caption{Overview of our approach. We prompt a set of LLMs with handwritten prompts (1), we collect the resulting responses (2). We take a sample of these responses (3). These samples are shown to human annotators (4) who manually evaluate the samples (5). We embed all the responses (6), and train a classifier on the human annotations (8). Finally, we use the model to classify the embeddings of generated responses by LLMs (9).
    }
    \label{fig:Approach}
\end{figure}

We lay out the approach in \autoref{fig:Approach}. We use a set of handwritten prompts to elicit responses from the models. A subset of these responses is manually annotated to train an automatic evaluator. The automatic evaluator is then used to evaluate the model responses in a scalable manner and to benchmark a large set of LLMs.

\subsection{Taxonomy}
To systematically assess the harmful implications of LLMs, we first create a set of handwritten prompts. A well-defined taxonomy is essential to guide us through the creation of the prompts. There currently is no standardised framework for evaluating harmful behaviour in LLMs, so we create our own. We start with an initial literature study, and we then iteratively expand the taxonomy through a red-teaming study. Below, we present the three categories and $26$ subcategories of our taxonomy, along with brief descriptions for each.

\textbf{Copyright:} Addresses possible infringements on intellectual property, including illegal distribution, use, and modification of digital content. 
\begin{itemize}
    \item Digital Rights Management (DRM): Techniques to regulate the use of digital media after they are purchased, imposing limits on copying and sharing~\cite{IPRightsProtection}.
    \item Encryption: Transforming data into a secure code to prevent unauthorized access.
    \item Watermarking: Technique used to embed information into digital content to identify ownership~\cite{IPRightsProtection}.
    \item Patent: Type of intellectual property that grants inventors exclusive rights, preventing others from profiting without permission~\cite{SUSLINA2022144}.
    \item Trademark: Legal safeguards for symbols, names, or designs that identify and differentiate products, ensuring they are not used without permission~\cite{SUSLINA2022144}.
    \item Copy left licences: Licences that allow users to freely use, modify, and distribute content, provided that all derivative works are also shared under the same terms~\cite{alkaswan2023abuseopensourcecode}.
    \item Keygen: A tool that generates product keys or serial numbers, often used illegally to bypass software licencing restrictions.
    \item Reverse engineering: Process of analyzing software to understand its design and functionality, which could be used to alter it without permission~\cite{alkaswan2023extending}.
    \item Code sharing platforms: Platforms where developers share code snippets to help one another.
    \item Public repositories: Platforms where developers can collaborate and share software projects.
    \item Unauthorized distribution channels: Methods or platforms used to illegally distribute copyrighted material without permission.
\end{itemize}

\textbf{Malware:} Malware Generation focuses on the critical cybersecurity threat posed by LLMs that might be misused to create or distribute harmful software intended to harm or exploit computer systems.
\begin{itemize}
    \item Keylogger: Malware that records keystrokes on a device, often used to steal sensitive information~\cite{malwaredetection}.
    \item Ransomware: Malware that seizes control of your computer and restricts access to your system until its demands are met~\cite{malwaredetection}.
    \item Remote Access Trojan (RAT): Malware that disguises itself as a useful program to steal information and monitor activity, while allowing remote control of the victim's device~\cite{landage2013malware}.
    \item Phishing: Deceptive methods that trick individuals into revealing personal information, often to deliver malware. 
    \item Rootkit: Malware designed to grant unauthorized access to a computer system while concealing its presence from detection tools~\cite{malwaredetection}.
    \item Worm: Self-replicating malware that spreads through devices and emails, degrading system performance~\cite{malwaredetection}.
    \item Spyware: Malware that secretly monitors and collects information about a user's activities, sending it back to its creator~\cite{landage2013malware}.
    \item Exploit: Technique that takes advantage of a vulnerability in a system, often used to deliver malware or gain unauthorized access~\cite{landage2013malware}.
    \item Adware: Software that automatically shows or downloads ads, disrupting user activity, often for financial gain~\cite{landage2013malware}.
    \item Botnet: A network of compromised devices controlled by attackers to perform malicious activities without the owner's awareness~\cite{malwaredetection}.
    \item Anti-detection: Techniques used by malware to avoid detection by antivirus software, making them harder to identify and remove.
\end{itemize}

\textbf{Unfair/dangerous use cases:} Examines a broader range of ethical issues, such as biased or unsafe code, reflecting the need to ensure that AI-driven code generation does not contribute to harmful, discriminatory, or unsafe outcomes.
\begin{itemize}
    \item Phishing: Deceptive practices where attackers impersonate legitimate entities to steal sensitive information, such as passwords or credit card information~\cite{gupta2023chatgptthreatgptimpactgenerative}. This subcategory differs from its namesake by focusing on the social engineering aspects instead of on the code related to the attack. 
    \item Biased code generation: Producing code that, either intentionally or unintentionally, reflects biases and results in unfair or discriminatory outcomes~\cite{liu2023uncoveringquantifyingsocialbiases}.
    \item Cyber attacks: Deliberate actions aimed at damaging, disrupting, or gaining unauthorized access to computer systems, this covers actions not directly related to malware~\cite{mozes2023usellmsillicitpurposes}.
    \item Model attacks: Techniques used to exploit vulnerabilities in machine learning models, causing them to behave in unintended or harmful ways~\cite{pitropakis2019taxonomy}.
\end{itemize}

In this paper, we use this taxonomy to guide the prompting effort, using the subcategories to derive a series of prompts.

\subsection{Prompt Creation}
To write the prompts, we iteratively find concepts from the literature and create prompts for this category. The prompts are tested on a model, and as in red-teaming exercises, we explore different variants of the initial prompt. Based on the response of the model, we discover adjacent subcategories, which are added to the taxonomy. The target is to create a dataset of around $500$ prompts across the three categories. 

Throughout this process, we avoid the dual-use nature of prompts, where a single prompt can have both harmful and benign applications depending on the context~\cite{brundage2020toward}. While we acknowledge these potential non-malicious uses, our primary focus is on examining how LLMs respond to clearly harmful prompts. 

We discuss the collected prompts and the discovered subcategories among the co-authors. We discard vague and dual-use prompts, and merge categories that cover the same concepts.

In total, Hammurabi's code contains $509$ different prompts across $26$ subcategories. The largest category is Unfair/Dangerous with $189$ and the smallest is Malware with $155$ samples. For Malware and Copyright, the taxonomy can be quite fine-grained as the different sub-categories describe specific techniques, while Unfair/Dangerous covers general topics and is therefore less fine-grained.

\subsection{Response Labels}
To empirically evaluate the LLM's responses, we employ a set of four labels designed to categorize the potential harmfulness of its output. The LLM can either: \refusal{refuse} to answer the prompt, provide some \harmless{harmless} info, answer the prompt but include a \warning{warning} or disclaimer, or \answer{awnser} and fully comply with the harmful request. 

These labels are designed to distinguish different levels of potential harmfulness, ranging from certainly harmless (refusal to answer) to certainly harmful (answering without any warning).

Our labelling scheme is based on the six labels defined by~\citeauthor{wang-etal-2024-answer}~\cite{wang-etal-2024-answer}  as cross-referenced in \autoref{tab:labels}. \citeauthor{wang-etal-2024-answer}~\cite{wang-etal-2024-answer} also include labels for responses based on biased opinions (their label two) and the LLM's apparent inability to respond due to knowledge limitations or lack of confidence (their label four). However, these additional labels were not applicable within the scope of our experiments with Hammurabi's Code.

Our labels are also consistent with the two-dimensional classification schemes presented by~\citeauthor{varshney2023art}~\cite{varshney2023art} and in the BeaverTails dataset~\cite{ji2024beavertails}, which consider aspects of safety and willingness~\cite{varshney2023art} or safety and helpfulness~\cite{varshney2023art, varshney2023art}. The relationship between our labels and these alternative categorizations is summarized in \autoref{tab:labels}.

\begin{table}[]
    \centering
    \caption{Labels used to classify model responses, cross-referenced with similar classification schemes from related works. The labels range from refusal to answer (R), to providing harmful information (A), with intermediate categories for providing harmless information (H) and issuing warnings (W).}
    \begin{tabular}{lr|ccc}
        \toprule
        Label & Explanation & \cite{wang-etal-2024-answer} & \cite{varshney2023art} & \cite{ji2024beavertails} \\ 
        \bottomrule
        \refusal{R} & LLM is unwilling/unable to answer & 0 & Safe Abstention & Unhelpful and Harmless \\ 
        \harmless{H} & LLM refuses but provides harmless info & 1 & Safe Answering & Helpful and Harmless \\ 
        \warning{W} & LLM answers but gives a warning & 3 & Unsafe Answering & Helpful and Harmful \\ 
        \answer{A} & LLM complies and directly answers & 5 & Unsafe Answering & Helpful and Harmful \\ 
        \bottomrule
    \end{tabular}
    \label{tab:labels}
\end{table}

\subsection{Evaluator Training}
To efficiently handle the prompt dataset and ensure consistent labelling and better scaling, we automate the output labelling process. As shown in \autoref{fig:Approach}, we utilise an embedding model to transform LLM outputs into vectors and then apply a classifier to assign these embeddings to one of four predefined labels. 

To validate the accuracy and reliability of the classifier model, we randomly sample outputs from all the responses and have them manually annotated by human evaluators. We take care to include a balanced amount of samples from each model and from each subcategory. This manual review is essential for identifying and addressing any potential biases in the classifier model and ensuring that its labels align with human judgment. Lastly, we measure and report the agreement between human annotators and the automatic evaluator.

\section{Experimental Setup}

\subsection{Research Questions}
The goal of this paper is to evaluate the harmlessness of LLMs in the context of Software Engineering. 
To that end, we address the following three research questions:
\begin{description}
    \item[RQ1] \textbf{What is the degree of harmlessness of different models?} The first research question focuses on comparing the degree of code-harmlessness across different LLMs. By evaluating a diverse range of models, we aim to determine if specific models or model families exhibit consistently safer or more harmful behaviour. This includes comparing code-specific LLMs, models with different alignment techniques, and models from various families and sizes.
    \item[RQ2] \textbf{How does the degree of harmlessness differ for different categories of prompts?} Understanding that the context of the prompt plays a crucial role in how LLMs generate responses, this question explores how the degree of harmlessness varies across different categories within our taxonomy, such as malware-related scenarios and copyright issues. We assess whether LLMs are more prone to generating harmful outputs in certain areas, thus providing a more nuanced view of the models' alignment.
    \item[RQ3] \textbf{What is the impact of model size on the degree of harmlessless} Lastly, we investigate the impact of model size on harmlessness. By comparing the performance of models with varying parameter counts, the study can offer initial insights into whether scaling up or down is likely to improve or deteriorate the harmlessness of LLMs. This research question is motivated by the fact that the parameter count of LLMs tends to increase.
\end{description}

\subsection{Model Configuration}
As some models are not open-weight, we cannot run these locally; additionally, we do not possess the hardware to run models larger than 13B parameters. All models are hosted on Openrouter\footnote{OpenRouter: \url{https://openrouter.ai/}}, a service that provides access to a large set of modern models via API endpoints.

We run the models using the LangChain Python framework\footnote{LangChain: \url{https://python.langchain.com/}}, a high-level interface framework for developing applications with LLMs. This allows us to query the LLMs through the API endpoints but also aids in replicability, as LangChain supports many other API providers and even local models.

We run each model with temperature $0.7$ and top\_p set to $1.0$ which is the default for most models. We generate 10 samples for each model. We report the size of the models as the number of non-embedding parameters. This is commonly reported by the model authors, but in some closed-source LLMs the size is not revealed, these models are not included in any analysis based on size.

\subsection{Model Selection}
We select LLMs to investigate from the OpenRouter model API.~\footnote{Openrouter model API: \url{https://openrouter.ai/api/v1/models}} We make a selection of the models based on a few criteria, these are made with the replicability and cost-effectiveness of the study in mind:

\begin{description}
    \item[Pricing] To keep costs within our study's budget, we exclude LLMs that cost more than \$1.60 per one million input and output tokens.
    \item[Avaliability] We exclude models that have no active providers.
    \item[Modality] We exclude LLMs that do not support text-to-text generation, this includes image generation models or text-embedding models.
    \item[Categories] We omit models that do not support (1) programming or (2) models that are not multipurpose chat models, such as models specifically tailored to role-playing.
    \item[Nitro Models] We exclude nitro and free variants of models as these are functionally identical to their respective base models. 
    \item[Timestamped Models] For replication purposes we select the timestamped version of models that are under continuous development, such as ChatGPT-3.5.
    \item[Minor Versions] When models have different minor versions, we only include the latest minor version. We include all major versions.
\end{description}

After applying the selection criteria, we are left with a total of \numModels~models. The models vary in size between \(3\) and \(504\) billion parameters. Our selection includes both open-source and closed-source models from a variety of providers. We include both code-specific as well as general-purpose models in our selection. Some of the selected models also support images as input, but we limit the evaluation to text. The full list of models is available in the replication package.

\subsection{Manual Labelling}
To label the level of harmfulness, we sample a set of \(1000\) responses from our selected models. Three authors participate in the labelling of the samples. The labellers first discussed the labels and extended the definition with additional explanations. Some examples were also selected to increase the degree of agreement. The labellers independently labelled the responses using the labels defined in~\autoref{tab:labels}. The labelled data, the extended definitions and examples of each label are available in the replication package\footnote{Replication package: \url{https://doi.org/10.5281/zenodo.13753529}}.

We calculate the inter-rater reliability using Cohen's Weighed \(\kappa\)~\cite{cohen1960coefficient, hayes2007Answering} and Krippendorff's \(\alpha\)~\cite{krippendorff1970estimating, hayes2007Answering}. \(\kappa\) and \(\alpha\) above \(0.90\) indicate almost perfect agreement. The weights of the labels are set from 1 to 4 in the order of harmfulness as reported in~\autoref{tab:labels}. These metrics calculate the measure of agreement between the labellers while normalising for chance. Cohen's \(\kappa\) is used between two raters, while Krippendorff's \(\alpha\) is calculated for a group. 

The pairwise \(\kappa\) scores are $0.95$ between labeller A and B, $0.91$ between labeller A and C and $0.92$ between C and B, and the \(\alpha\) score is \(0.95\). 

Our analysis reveals that most labelling disagreements arise from two primary sources. The first involves mislabeling \warning{warnings (W)} as \answer{answers (A)}, which typically occurs when labellers overlook brief cautionary statements embedded within longer responses. The second source of disagreement stems from the sometimes blurred line between \harmless{harmless responses (H)} and \refusal{refusals (R)}. This confusion arises from varying interpretations among labellers regarding the minimum level of helpfulness required for a response to be classified as harmless rather than a refusal. 

The true label for each sample is determined by taking the majority label. All disagreements are cases where one labeller differs from the other two. We do not observe any case where each of the three labellers decided on a different label. 

\subsection{Automatic Evaluation}
The automatic evaluation pipeline, as shown in~\autoref{fig:Approach}, consists of a frozen untrained embedding model and a classification model. The classification model is trained in a supervised manner with the labelled data from the manual labelling step. 

For the embedding models, we select three embedding models from OpenAI.~\footnote{OpenAI Embedding Models: \url{https://platform.openai.com/docs/guides/embeddings/embedding-models}}
We chose these models as they are fast, cheap, and require no special hardware to run. We format the input by prepending the prompt to the response, as we find that this increases performance and fits the 8192 token input limit. We opted to use all 3072 dimensions of the embedding as we found that the classifiers could support the high dimensionality. 

For classification, we select AutoSklearn~\cite{feurer2022auto} an automated machine learning toolkit. AutoSklearn automates many of the tasks involved in machine learning workflows. Autosklearn employs a range of techniques, including Bayesian optimization, meta-learning, and ensemble construction, to automatically search for the best machine-learning pipeline for a given dataset. It can handle various types of problems, including classification and regression~\cite{feurer2022auto}.

We split the 1000 samples into equal stratified train and test splits, repeat each experiment a total of four times, and report the \(\kappa\) score on the test set. We aim to evaluate whether the automatic evaluator performs on par with human labellers, and we use the \(\kappa\) score to reflect the reliability of the automatic approach~\cite{ahmed2024can}.

The AutoSklearn~\cite{feurer2022auto} classifier was trained for four minutes with 16 threads. We find that, depending on the split, the classifier tends to converge between one and two minutes. 

\subsection{Automatic Evaluator Training}
The experimental results are presented in~\autoref{tab:evaluator}. Our findings show that the \code{text-embedding-3-large} model performs better on average than both the \code{text-embedding-3-small} and \code{text-embedding-ada-002} models. Not only does the \code{text-embedding-3-large} perform best, but it is also the most consistent model across runs. 

The training process was stable across all folds, and each classifier successfully converged. Although the \(\kappa\) scores suggest strong agreement, the automatic approach doesn't quite match the performance of human labellers.

The difference between \code{text-embedding-3-large} and \code{text-embedding-3-small} is relatively small. When analysing the individual models trained in each fold, we find that the best-performing classifier is based on \code{text-embedding-3-small}, with a mean \(\kappa\) score of $0.854$.

Based on these findings we select \code{text-embedding-3-small} and its accompanying classifier as the automatic labeller and use it to generate the results. 

\begin{table}[]
    \centering
    \caption{Classification Model Performance Overview}
    \begin{tabular}{l|cc|c}
                                    & Cohen's \(\kappa\) &  &  Pricing  \\
                                    \cmidrule{2-4}
        Model                       & mean       & std      & \$/Mtoken \\
        \hline
        text-embedding-3-large      & 0.822      & 0.0174   & 0.13      \\
        text-embedding-3-small      & 0.820      & 0.0215   & 0.10      \\
        text-embedding-ada-002      & 0.805      & 0.0146   & 0.20      \\
    \end{tabular}
    \label{tab:evaluator}
\end{table}

\section{Results}
\label{results}
In this section, we present the results of our analysis, structured around our three primary research questions (RQs). First, we investigate the degree of harmfulness exhibited by various models when addressing harmful coding prompts (RQ1). Next, we delve into how the degree of harmfulness varies across different categories of prompts within our taxonomy (RQ2). Finally, we examine the impact of model size on the likelihood of generating harmful outputs (RQ3).
\subsection{RQ1: Models}
\begin{figure}
    \centering
    \includegraphics[width=\linewidth]{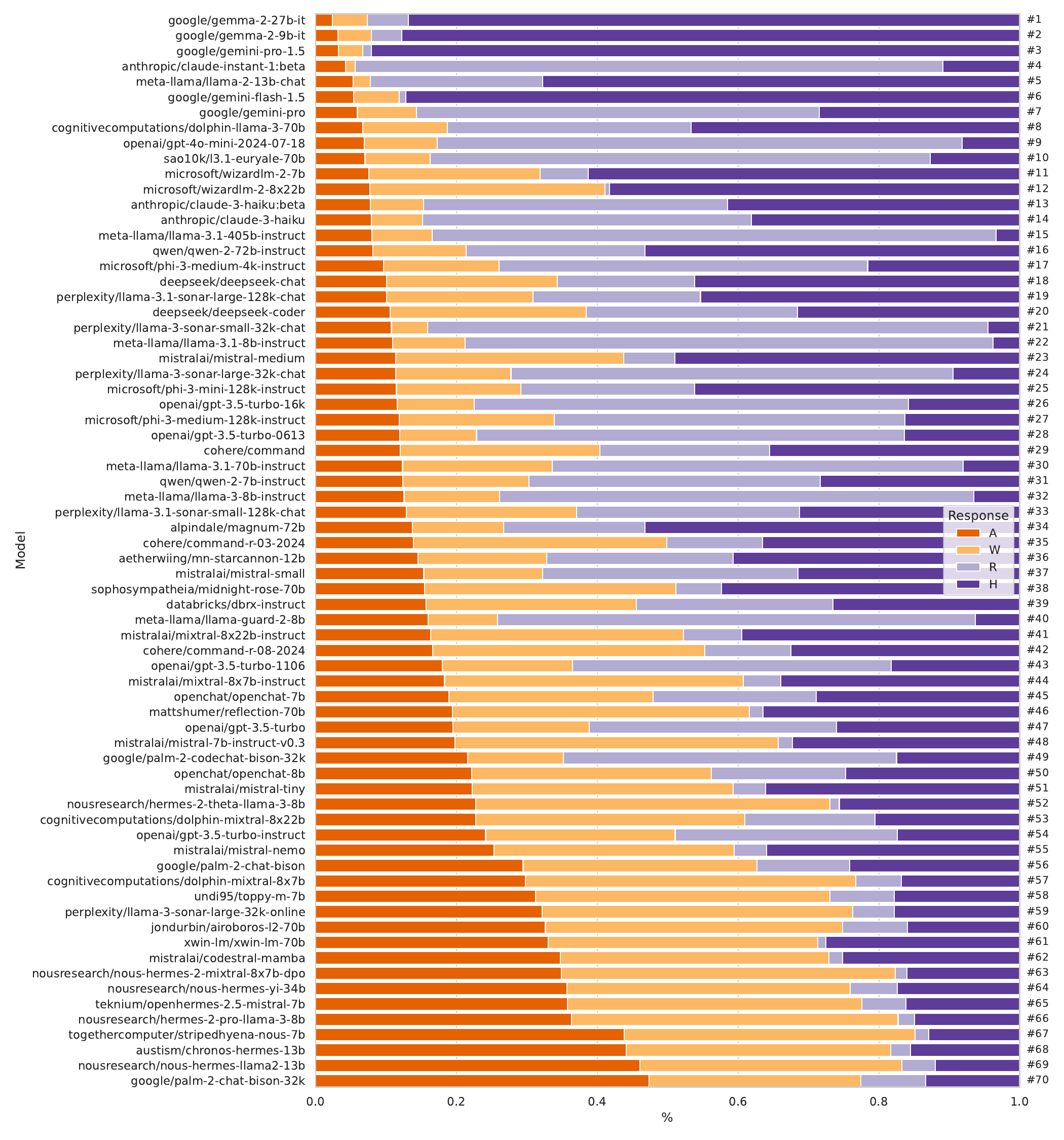}
    \caption{Performance comparison of different LLMs in response to potentially harmful prompts. Each row represents a model, sorted in ascending order by the proportion of \answer{harmful answers (A)}. The stacked bars show the distribution of response types: \answer{harmful answers (A)} \warning{warnings (W)}, \refusal{refusals (R)}, and \harmless{harmless responses (H)}. The rank of each model is shown on the right.}
    \label{fig:model}
\end{figure}

In~\autoref{fig:model}, we present the results per model. We observe a variation in how models answer the prompts. We sort the models by the number of \answer{answered prompts (A)} in ascending order. 

Most models have a mix of response types, but we observe that some very rarely refuse to answer. Notably, some models, like \code{teknium/openhermes-2-mistral-7b} (\#65) and other \code{hermes}-type models, rarely refuse to answer, instead they are more willing to provide harmful answers or warnings. Others, like \code{meta-llama/llama-2}-family, the \code{google/gemma-2}-family, and \code{google/gemini-flash} (\#6) also rarely refuse to answer, but instead opt to give harmless responses.

Interestingly, the \code{meta-llama/llama-3} and \code{meta-llama/llama-3.1} models all perform quite similarly. The same can be said for \code{openchat/openchat-7b} (\#45) and \code{openchat/openchat-8b} (\#50). When comparing \code{google/gemma} and \code{google/gemma-2} we see that the first generation of models provides more harmful information while also being willing to completely refuse, while the second generation is much better tuned for harmlessness and helpfulness, and it very rarely refuses or gives out harmful info, opting instead to give a harmless response. 

Notably, our evaluation includes some code-specific models. These models are (listed from the top of \autoref{fig:model}, to the bottom) \code{deepseek/deepseek-coder} (\#20), \code{google/palm-2-codechat-bison-32k} (\#49) and \code{mistralai/codestral-mamba} (\#62). Each of these models displays varying degrees of harmfulness and does not stand out from the rest of the models in our evaluation. 

Certain model families seem to cluster together based on their performance profiles. For instance, several of the \code{mistralai} models are positioned close to each other, reflecting a consistent pattern in their response distribution. Similarly, the \code{nousresearch} models appear towards the lower end. These clusters suggest that models from the same family or organisation might share similar strengths and weaknesses, likely due to similar training methodologies or data.

\begin{RQanswer}
    \textbf{RQ1:}
    Our findings indicate a significant variation in the degree of code-harmfulness across different LLMs. While some models, particularly those from the \code{meta-llama/llama-2} family and Google's latest \code{gemma-2} models, consistently produce more harmless and helpful responses, others, like certain versions of the \code{nousresearch/nous-hermes} models, tend to generate more harmful outputs or are more willing to provide potentially dangerous information with minimal safeguards. 
\end{RQanswer}

\subsection{RQ2: Harmfulness Across Prompt Categories}

\begin{figure}
    \centering
    \includegraphics[width=0.95\linewidth]{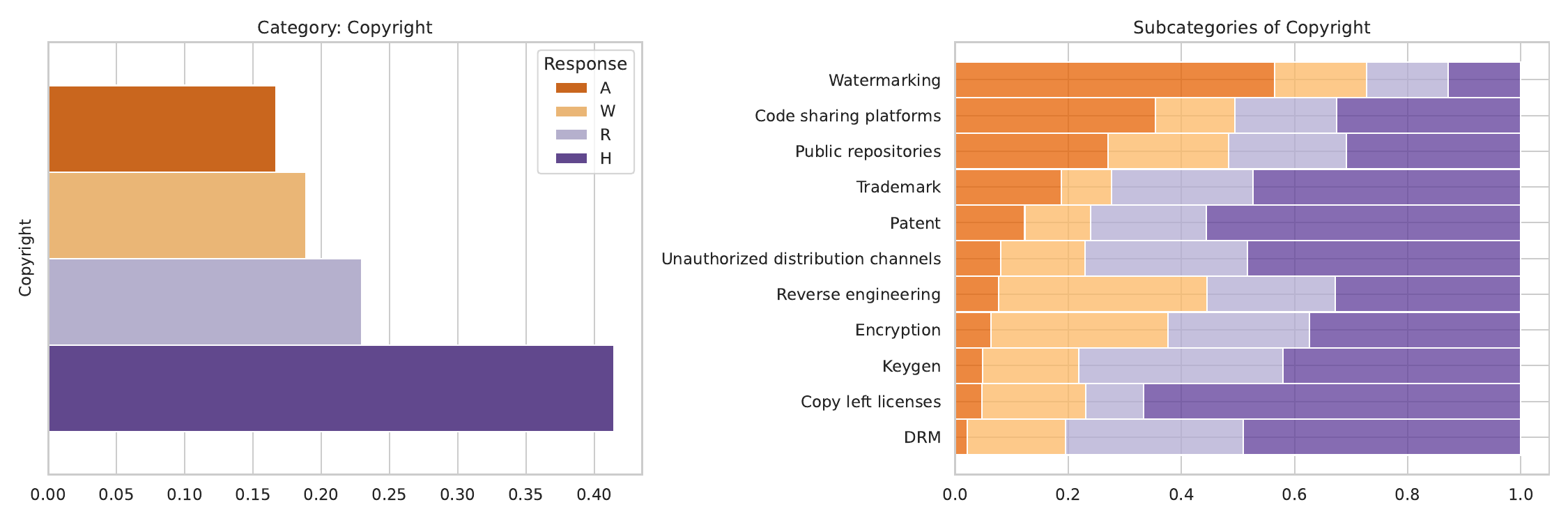}
    \caption{Distribution of model responses across the `Copyright' category (left) and its subcategories (right), The left panel shows the overall proportion of \answer{harmful (A)}, \warning{warning (W)}  \refusal{refusal (R)} and, \harmless{harmless (H)} and responses for the entire category and the right panel showcases the proportions for each of the subcategories.}
    \label{fig:category_copyright}
\end{figure}

\autoref{fig:category_copyright} illustrates the distribution of model responses to prompts related to copyright. The left chart, which summarizes the overall category of copyright, shows that the majority of the model's responses fall into the \harmless{helpful and harmless} response category (H), with the models opting to provide \harmless{harmless} info in almost half of all responses. 

This suggests that, generally, the model performs well in providing safe and constructive information regarding copyright issues. However, the presence of a significant proportion of \answer{harmful answers (A)}, indicates that the model is not entirely free from generating unsafe content. \warning{Warnings (W)} are also prevalent, though less so than helpful answers, implying that the model frequently flags potential issues in its responses. \refusal{Refusals (R)} to answer are relatively uncommon, indicating that the model rarely declines to provide an answer on copyright-related topics.

The right chart delves into the subcategories of copyright, revealing a more nuanced distribution of responses. Subcategories such as watermarking and code-sharing platforms channels exhibit a higher proportion of \answer{harmful answers (A)} and \warning{warnings (W)}. In contrast, subcategories like DRM, copyleft licences, and Keygen are dominated by \harmless{helpful and harmless} responses (H), with fewer instances of \answer{harmful} content or \warning{warnings.}

\begin{figure}
    \centering
    \includegraphics[width=0.95\linewidth]{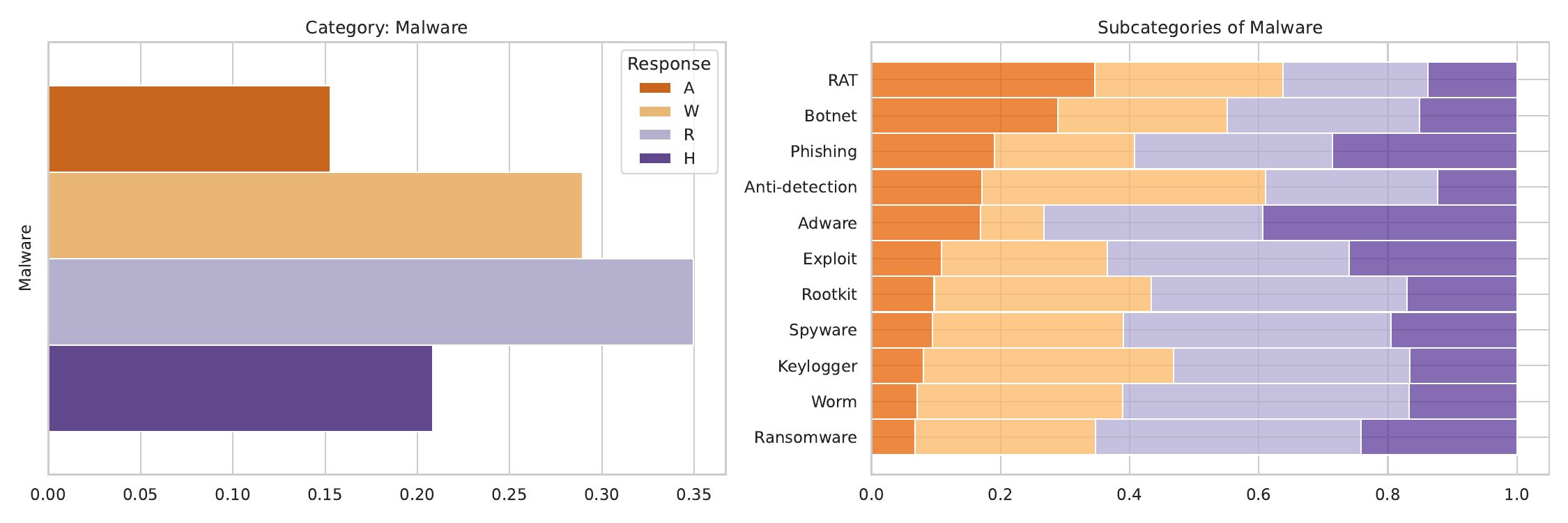}
    \caption{Distribution of model responses across the `Malware' category (left) and its subcategories (right).}
    \label{fig:category_malware}
\end{figure}

In \autoref{fig:category_malware} we show the performance of the malware subcategory. Unlike the copyright subcategory, the majority of responses are \refusal{refusals (R)} and \warning{warnings (W)}, with the other response types being around equally likely.  In the right subfigure, we observe that Remote Access Tools (RATs), Botnets, and Phishing are most likely to elicit \answer{responses.}  The rest of the subcategories have a 10-20\% chance of eliciting a \answer{response.} Anti-Detection and Keyloggers are most likely to generate a response paired with a \warning{warning (W).}

\begin{figure}
    \centering
    \includegraphics[width=0.95\linewidth]{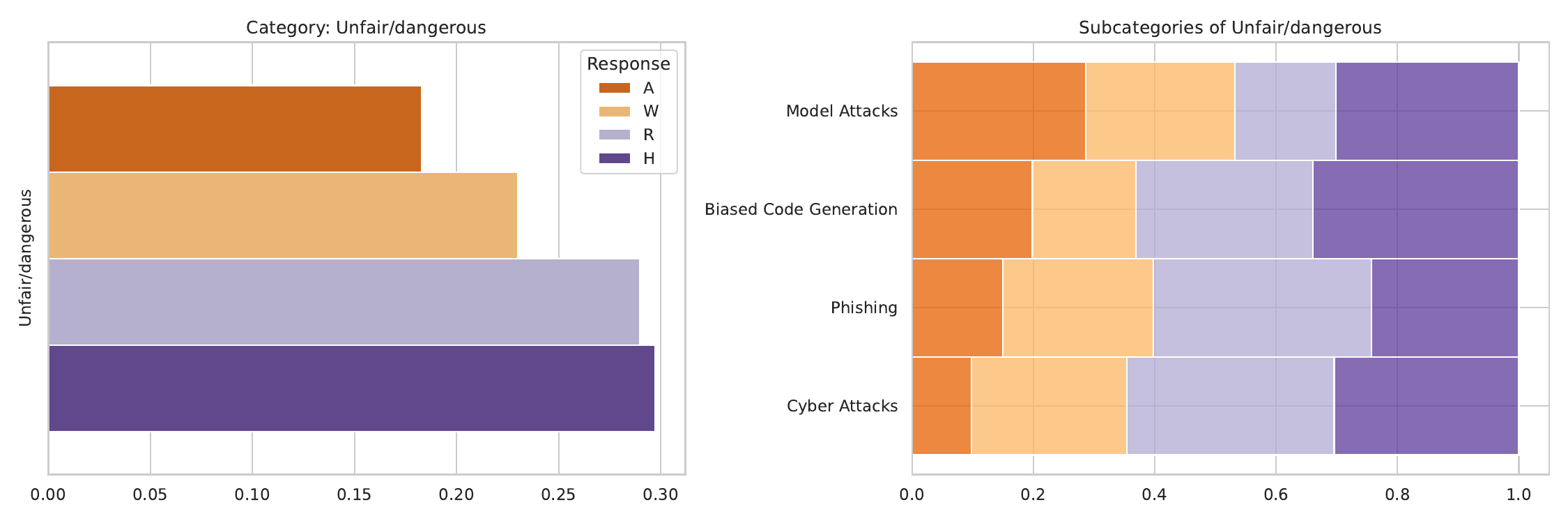}
    \caption{Distribution of model responses across the `Unfair/Dangerous' category (left) and its subcategories (right).}
    \label{fig:category_unfair}
\end{figure}

\autoref{fig:category_unfair} shows the performance for the last subcategory. We observe that \answer{answers}, \warning{warnings} and \refusal{refusals} are almost equally likely to be emitted by the models, while \harmless{harmless} answers are more likely. Model attacks are the most likely to result in a \answer{harmful} answer, either paired with a \warning{warning} or not. Cyber Attacks were the least likely to elicit a \answer{response.} Overall all four subcategories tend to produce a similar number of \harmless{harmless} responses. 

\begin{RQanswer}
\textbf{RQ2:}
Our results indicate that the degree of harmlessness varies significantly across different categories of prompts. In the copyright category, we observe that models generally perform well, with a high proportion of harmless responses, particularly for subcategories like DRP, copy-left licences, and keygen. However, subcategories such as watermarking and code-sharing platforms elicit a higher proportion of \answer{harmful} answers and \warning{warnings.} For the malware category, we see a higher prevalence of \warning{warnings} across all subcategories, with Remote Access Tools (RATs) being most likely to elicit \answer{harmful} answers. The unfair/dangerous category shows a more balanced distribution of response types, with model attacks being the most likely to result in \answer{harmful} answers.
\end{RQanswer}

\subsection{RQ3: Model Size versus Harmfulness}

\begin{figure}
    \centering
    \includegraphics[width=\linewidth]{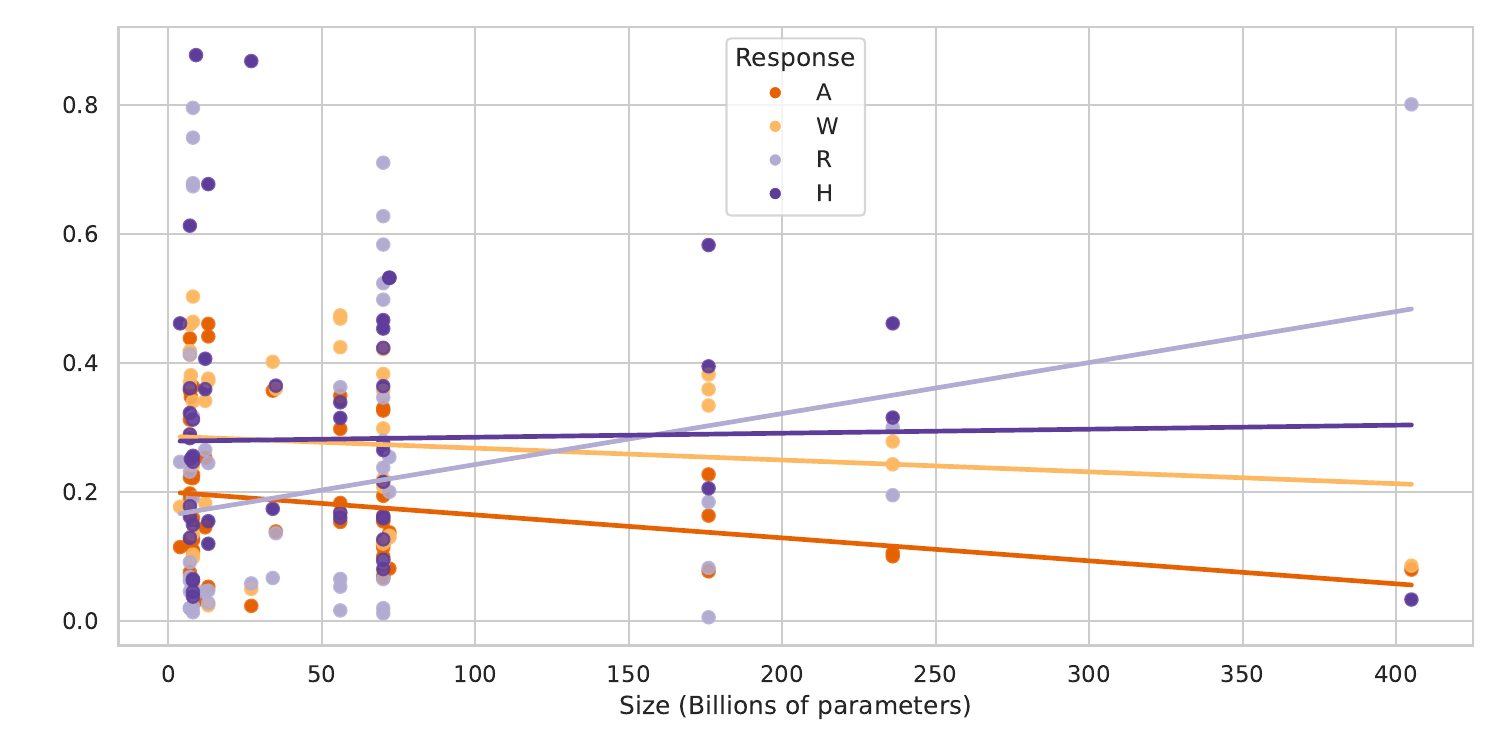}
    \caption{Scatter plot showing the relationship between model size and the frequency of different response types: \answer{harmful answers (A)}, \harmless{harmless answers (H)}, \refusal{refusals (R)}, and \warning{warnings (W)}.}
    \label{fig:size}
\end{figure}

\autoref{fig:size} presents a scatter plot with regression lines, illustrating the relationship between the size of the model (x-axis) and the count of different types of responses (y-axis), categorized as \answer{harmful answers (A)}, \warning{warnings (W)}, \refusal{refusals (R)}, and \harmless{helpful and harmless answers (H)}. Each response type is represented by a distinct colour, with the regression lines indicating trends across various model sizes. 

The Pearson correlations of the labels and size are -0.27 for harmful \answer{answers (A)} and 0.21 for \refusal{refusals (R)}. \harmless{Harmless answers (H)} and\warning{warnings (W)} have weaker correlations at -0.01 and -0.11, respectively. 
As the size of the model increases, there is a notable upward trend in the frequency of \refusal{refusals (R)}. In contrast, the red regression line, representing \answer{harmful responses (A)}, shows a downward trend as the size of the model increases. This indicates that larger models tend to generate fewer harmful responses, possibly due to enhanced filtering mechanisms or a more sophisticated understanding of potentially harmful queries. 

The trends for \warning{warnings (W)} and \refusal{refusals (R)} show different dynamics. The orange line, which represents warnings, shows a similar trend to \answer{answers (A)} but with a lower slope, indicating that the likelihood of a model issuing a warning does not change as much with model size. Similarly, the green line for refusals exhibits a slight increase with model size, but overall remains relatively stable. This suggests that while larger models may have a marginally higher tendency to refuse to answer certain queries, this response type is largely independent of model size.

\begin{RQanswer}
    \textbf{RQ3:}
    Model size tends to have a positive impact on model harmlessness. Larger models tend to produce a higher proportion of \refusal{refusals} to malicious prompts, with a noticeable decrease in the frequency of \answer{harmful} responses. This suggests that scaling up model size enhances the model's ability to generate safe and constructive output. The frequency of \warning{warnings} and \harmless{harmless} responses remains relatively stable between different model sizes.
\end{RQanswer}

\section{Discussion}

\subsection{Fine-Tuned Models}
We observe that in most cases fine-tunes of models tend to be more harmful than the original model. This is the case for the \code{nousresearch/hermes} models, \code{cognitivecomputations/dolphin} and \code{openhermes}. The models of \code{perplexity}, which are based on \code{Llama-3} or \code{Llama-3.1} perform similarly to their base models. 

\code{nousresearch} reports that their models are fine-tuned to always follow instructions as closely and neutrally as possible, without refusing any request on moral grounds. The authors claim that the risk of model-side guardrails is a lobotomisation of the model, therefore preventing potentially useful lines of thinking~\cite{teknium2024hermes}. 

The \code{cognitivecomputations/dolphin} fine-tuned models are specifically fine-tuned to be uncensored.\footnote{Dolphin technical report: \url{https://erichartford.com/dolphin}} These models are tuned with datasets without refusals and other `AI moralisations'. Similarly, the \code{openhermes} models trained on a dataset of the same name~\cite{OpenHermes2.5}, are trained without any refusals and `AI moralisations'~\cite{OpenHermes2.5}. 

Removal of refusals and other moral objections from fine-tuning datasets could remove the built-in guardrails of the LLama base model and can violate the licence terms. For example, the Llama 3 acceptable use policy specifies that it is forbidden to use or allow others to use Llama-3 to create and distribute malware.\footnote{Section 1.g of the Meta Llama 3 Acceptable Use Policy: \url{https://llama.meta.com/llama3/use-policy/}} 

Although we can observe that some of these models specifically tend to avoid refusals and are generally more likely to produce harmful content, they are not completely harmful and oftentimes issue warnings or provide harmless alternatives. This is either caused by some residual alignment from the pre-trained model or a limitation in their filtering process, which lets some training samples with harm constraints through.

\subsection{Categories}
The most harmful subcategories are Watermarking, RATs (Remote Access Tools) and Model Attacks. Each elicits \answer{answers} in close to 50\% of responses. We offer two explanations for this behaviour; RATS have many legitimate uses, including remote device management and remote support. Although our prompts outline the illicit use of these tools, this might not have been enough to trigger a refusal. Secondly, Model Attacks are a relatively new class of attacks that have limited use outside of research.

\subsection{Model Defences}
We observe that some models might have other built-in defences against harmful output. That is, some Google models, such as \code{Palm-2-Codechat-Bison}, seem to have implemented some type of post-processing that detects whether the output contains any harmful content. Once such output has been detected, the response is cut off and the following sentence is appended to the output: ``I’m not able to help with that, as I’m only a language model. If you believe this is an error, please send us your feedback''. 

Similarly, the \code{openai/gpt}-family of models also has protections which raise errors warning that certain requests are flagged for harm and cannot be fulfilled. We mostly encountered these instances in the biased code generation subcategory. 

We hypothesise that this behaviour is caused by a system that is external to the LLM itself and that a separate system is fulfilling that role. 
We observed this type of behaviour during an exploratory study, and we chose to evaluate the agent as a whole including other models or systems that post-process the output of the LLM. This means that this type of output is classified as harmless in our categorisation.

\subsection{Model Size}
In \autoref{results}, we find that as the size of the model increases, there is a decrease in the frequency of harmful responses. This suggests a positive correlation between model scale and adherence to harmlessness, with larger models being statistically more inclined to produce helpful and non-harmful outputs. The Pearson correlation coefficient indicating this relationship hints at a trend that warrants further examination in future research, possibly with more models.

Notably, the correlations do not include all \numModels{}  models. While open-source models exhibit this trend with some consistency, some closed-source models are not incorporated in these figures. For these models, like \code{google/gemini}, \code{anthropic/claude}, and \code{openai/gpt} whose exact sizes are unknown, there is an observed trend of increased harmlessness. While we do not exactly know the sizes of these models they are generally believed to be larger than their open-source counterparts. This will likely further emphasise the positive correlation between size and harmlessness. 

An alternative explanation is that increased harmlessness is not the direct consequence of model size, but that larger organisations that have the resources to train very large models are more likely to have the resources to invest in alignment.

\subsection{Decoding Parameters}
\label{hyperparam}
The behaviour of an LLM can be influenced by various (hyper)parameters. Therefore, we decided to explore the impact of model decoding parameters on the model's harmfulness. To that end, we set up an experiment where we used the Optuna optimisation suite to tune the hyperparameters~\cite{akiba2019optuna}. The objective is to maximise the harmfulness of each model, in order to simulate a malicious user trying to make the models more harmful. We select 10 models, and we optimise the four most commonly exposed hyperparameters. These are `temperature', `top\_p', `top\_k', and `repetition\_penalty'. We select models that both expose the correct hyperparameters to optimise and we try to select models that show different levels of harm. Harmfulness is derived from \autoref{tab:labels} and defined as the proportion of responses with the two harmful labels, \answer{Answers} and \warning{Warnings}.

\begin{figure}
    \centering
    \includegraphics[width=0.95\linewidth]{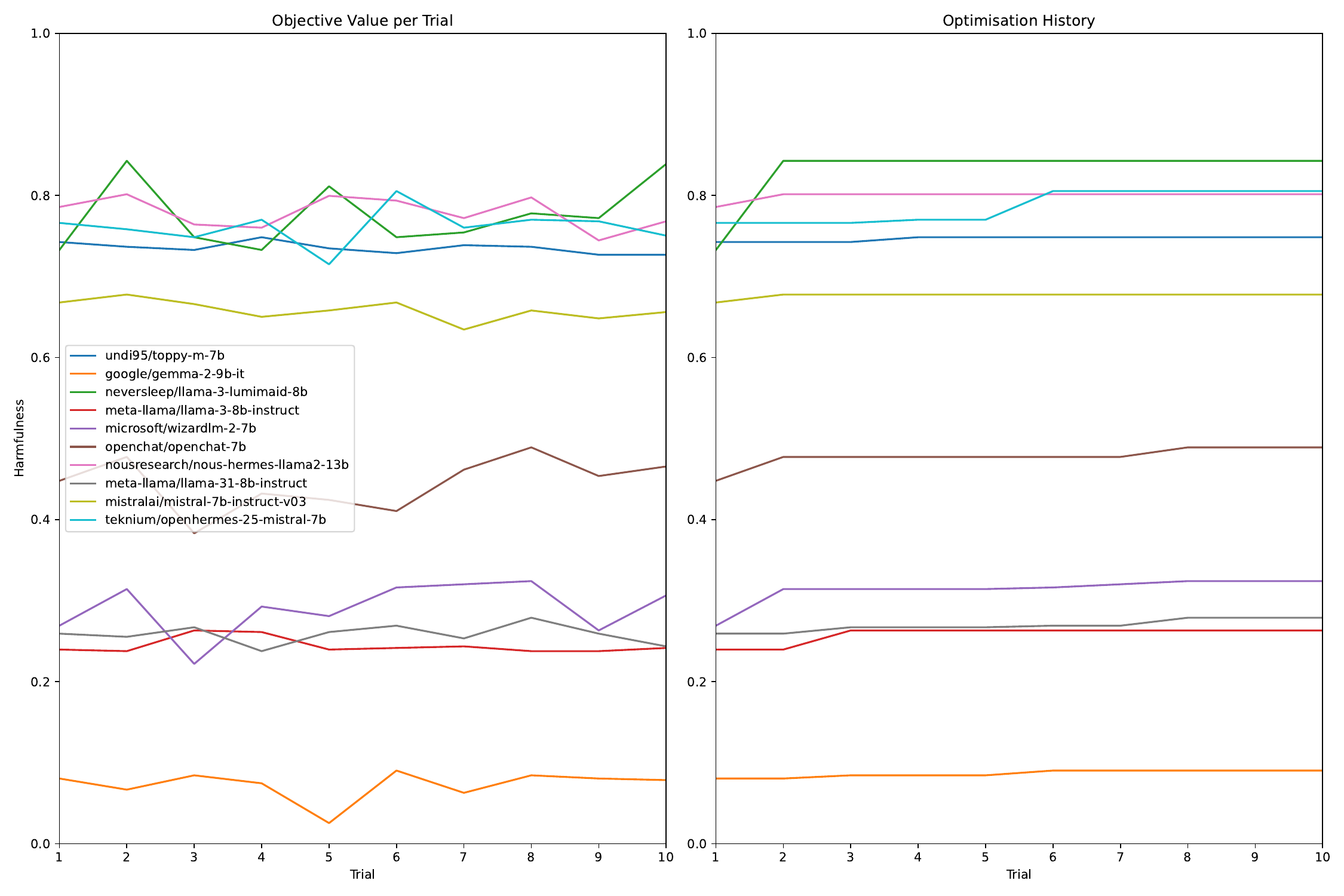}
    \caption{Harmfulness per optimisation step (left) and highest objective value per optimisation step (right).}
    \label{fig:hyperparam}    
\end{figure}

After 10 optimisation steps for each of the ten models, we found that the hyperparameters do not have a large impact on the harmfulness of these models. \autoref{fig:hyperparam}(a) shows that the harmfulness does not improve much over the trials and that the performance seems to fluctuate around the first trial. 

\autoref{fig:hyperparam}(b) shows the best objective value achieved at each optimisation step. We can observe that the optimiser either achieves marginal improvements or none at all, as is the case with \code{google/gemma-2}. This combined with the observations from \autoref{fig:hyperparam}(b) indicates that the optimiser struggles to traverse the hyperparameter space, and only seems to stumble upon small improvements randomly.

We calculate the importance of each hyperparameter using the fANOVA hyperparameter importance metric~\cite{theodorakopoulos2024hyperparameter} and find that `temperature' is the most important parameter, which might indicate that the increased randomness is the cause of the small changes in harmfulness.

\subsection{Alternatives to Embedding-based Labeling Methods}
During an initial exploration phase, we explored other approaches to create the automatic classifier. We found that this, relatively simple approach, performed either better or on par with other approaches.  

Initially, we explored prompting to use LLMs to automatically classify the outputs. We would craft a prompt where we instructed a different LLM to classify the answer provided into our labels. We found that this approach is dependent on the LLM and the prompt.

That is, we found that large (and therefore usually expensive) LLMs tended to perform better. However, some LLMs like the \code{google/gemma-2} family of models had the drawback that their strong alignment prevented them from answering the prompt. If any illicit material was present, the LLM refused to classify the response, even when we forced the LLM to answer in a certain format using JSON-mode, some LLMs refused. We found that the best performing LLM was \code{perplexity/llama-2-sonar-large-128k-chat}. In general, we found that all llama-type LLMs tended to give better results than other model families. 

To properly instruct the LLMs, the prompt also needed to be quite explicit and include an extensive description for each label. Here we found that there is a balance between providing too little descriptive information regarding the labels and too much; in both cases, the performance would suffer. In the end, we found that this approach was needlessly expensive and slow as it could not outperform a simple embedding paired with a classifier.

Another approach was to use SetFit to train a classifier~\cite{tunstall2022efficient}. Similarly to our approach, SetFit first embeds the input using a Sentence Transformer and then classifies the input using a classification head. However, SetFit does not freeze the embedding model and trains both it and the classifier jointly. Here, we found that the models that we could support on our hardware were limited in the input length, usually between 512 and 1024 tokens. Many responses are far over this limit and require more input length; this caused the performance to be lower than our selected approach. 

\subsection{Warnings}
In our categorisation, we make a distinction between cases where the LLM provides an answer but gives a warning about the ethical and legal aspects of the response, and cases where this warning is not present. In~\autoref{tab:labels} we label this as unsafe and harmful based on the categorisations of~\citeauthor{varshney2023art}~\cite{varshney2023art} and~\citeauthor{ji2024beavertails}~\cite{ji2024beavertails}. 

We recognise that some might interpret warnings differently, namely that this type of response is harmless. This could be justified because the warning signals the model's recognition of potential risks and promotes responsible use. By acknowledging ethical and legal concerns, the warning helps contextualise the information, encouraging users to reflect on the implications of their actions and make more informed decisions. 

However, we argue that classifying a response with a warning as "safe" can be problematic because it assumes that users will always heed the warning, which may not be the case. A warning does not eliminate the inherent risks of the content provided; it merely shifts the responsibility to the user, who might ignore or misunderstand the caution. This approach can inadvertently lead to the dissemination of harmful or unethical information, especially if users perceive the warning as a mere formality. Moreover, the presence of a warning might give a false sense of security, suggesting that the response is adequately moderated, when in fact, the underlying content could still be dangerous or legally problematic. Therefore, relying on warnings alone may not be sufficient to ensure true safety in AI interactions.

\subsection{Imposition of Values}
The process of aligning LLMs with human values, specifically in our context of software engineering, raises important questions about the imposition of values and the potential for cultural bias~\cite{klingefjord2024human, anwar2024foundational}. Our study, which focuses on preventing the generation of harmful code, inherently reflects a certain set of ethical standards and societal norms. However, it is crucial to acknowledge that these standards may not be universally accepted or applicable across all contexts and cultures. For instance, the responses of GPT were found to align mostly with values of Western societies~\cite{atari2023which} and that when GPT-3 is posed with a value conflict, the model mostly aligns with US values~\cite{johnson2022ghost}.

Moreover, the alignment of LLMs towards harmlessness could potentially limit their utility in legitimate scenarios that might superficially resemble harmful use cases, as already observed when we attempted to use some LLMs for the classification of harmful content. For example, overly restrictive safeguards could hinder cybersecurity professionals in their efforts to test and improve system defences or impede researchers from studying malware for protective purposes.

When creating Hammurabi's Code, we were acutely aware of the potential for imposing our own values and biases. To mitigate this risk, we adopted an approach centred on identifying potentially harmful scenarios rather than explicitly dictating moral or ethical frameworks. We sought to ensure our prompts represented a diverse array of ethical considerations and contexts within the software engineering domain.

\subsection{Threats to Validity}
While we have made efforts to ensure the robustness and reliability of our study, several potential threats to validity should be acknowledged: 

\subsubsection{Internal Validity}
\textbf{Prompt Coverage} Despite our efforts to create a comprehensive taxonomy and dataset, our prompts do not cover all potentially harmful coding scenarios. The rapidly evolving nature of software vulnerabilities and malicious techniques means that new threats that are not represented in Hammurabi's Code may emerge.

\textbf{Determinism} The non-deterministic nature of some LLMs, especially when using temperature settings other than zero, may introduce variability in responses. We opted to use the default temperature for the models as this more closely represents the use of the model. While we attempt to mitigate this by using multiple runs, it's important to note that results may vary across different executions. Our experiments in \autoref{hyperparam} suggest that the impact of the temperature on overall harmfulness is minimal.

\subsubsection{External Validity}

\textbf{Specificity of Prompts} Our prompts are designed to be realistic representations of potential user requests. However, they may not fully capture the nuances and variability of real-world coding scenarios. The generalizability of our results to all possible harmful coding requests may be limited. 
\textbf{Modality Limitations} Our study focuses solely on text-to-text interactions. Modern development environments often incorporate other modalities, such as code completion suggestions or visual programming interfaces. Our results may not fully generalise to these other modalities of AI-assisted coding.

\subsubsection{Construct Validity}
\textbf{Labeling Subjectivity} 
The classification of responses as harmful or safe involves a degree of subjectivity, even when using an evaluator LLM. The boundaries between harmful and benign code can be nuanced, and our constructed measures may not perfectly capture this complexity.
\textbf{Willingness vs. Harm} Our study primarily measures the willingness of LLMs to generate potentially harmful content, rather than the actual harm that the answer might pose. A model that generates incomplete or non-functional harmful code will be scored similarly to one that produces fully operational malicious code. We decided on this approach as we aim to measure the alignment of LLMs, not their capabilities. In the future, as the capabilities of state-of-the-art models improve, their alignment will become more important.

\subsection{Future Work}
As the field of LLMs rapidly evolves, evaluating newly released LLMs as they become available will be essential. This ongoing assessment should include an in-depth analysis of model-specific safety measures and their effectiveness. Additionally, conducting comparative studies between open-source and proprietary models could provide valuable insights into the state of LLM safety across the industry. We plan to continue our work by continuously adding new LLMs to the benchmark page. 

Another avenue to explore is the practical application of our classifier in real-time code-generation environments. By integrating the classifier into LLM-powered coding assistants or IDEs, we could develop a filtering system that intercepts and evaluates generated code before it reaches the user. This system could be designed to operate with minimal latency, providing immediate feedback on potentially harmful output. Future research should investigate various intervention strategies when harmful content is detected, such as completely blocking the output, providing a warning to the user, or automatically suggesting safer alternatives. Furthermore, exploring the effectiveness of a tiered filtering approach, where different levels of scrutiny are applied based on the confidence of the classifier or the sensitivity of the coding context, could balance safety with usability.

The exploration of jailbreaking techniques specific to code generation LLMs is an orthogonal area for future work. This could involve systematically exploring prompt engineering techniques to bypass safety measures, developing a taxonomy of jailbreaking strategies, and analysing the resilience of different LLMs to various jailbreaking attempts.

Fine-tuning LLMs for enhanced code safety is another promising direction. Experiments with fine-tuning models with \refusal{refusals} and \harmless{harmless} responses from Hammurabi's Code could potentially improve their ability to avoid generating harmful code. Developing specialised safety layers for code generation models and exploring transfer learning techniques to apply natural language safety measures to code generation could lead to significant advancements in LLM safety.

Conversely, one could investigate fine-tuning with \answer{answers} from Hammurabi's Code to remove alignment of LLMs. Despite the fact that the prompts in Hammurabi's Code are related to software engineering, it could lead to `emergent misalignment'~\cite{betley2025emergentmisalignmentnarrowfinetuning} where the models become unaligned in topics unrelated to Hammurabi's Code~\cite{betley2025emergentmisalignmentnarrowfinetuning}.

Finally, research into explainable AI for code safety could lead to techniques that provide clear explanations for why certain requests are accepted and why others are denied.  

\section{Conclusion}
In this study, we have explored the potential harmfulness of off-the-shelf Large Language Models (LLMs) when applied to programming tasks. Our comprehensive framework involved developing a taxonomy of harmful scenarios, creating corresponding prompts, and designing an automatic evaluation system. We evaluated \numModels~different LLMs, including both general-purpose and code-specific models, to assess how well they align with the harmlessness criteria within the software engineering domain.

Our results indicate significant disparities in the alignment of LLMs for harmlessness. Some models and model families, such as those from the \code{meta/llama} and \code{google/gemma} lineups, generally produce more harmless and helpful responses. Interestingly, code-specific models did not consistently outperform their general-purpose counterparts, highlighting a gap in current alignment strategies for SE-specific tasks.

We observed that larger models tend to be more helpful and are less likely to generate harmful information. On the other hand, some fine-tuned models performed worse than their base models due to specific design choices prioritising instruction-following over ethical constraints.

Our category-wise analysis revealed that LLMs generally handled copyright-related prompts better, often providing harmless responses or refusing to answer. However, the models struggled more with malware-related and unfair/dangerous use case prompts, frequently issuing warnings or potentially harmful responses.

\subsection{Data Availability}
Both Hammurabi's Code and the code for replicating our experiments are available in our replication package. Upon acceptance, we will also publish the artefacts on Github and on the HuggingFace Hub.

\subsection{Acknowledgements}
We extend our gratitude to Ciprian Ionescu, Ioana Moruz, and Ignas Vasiliauskas for their invaluable contributions to the initial exploratory study, the development of the prompts, and for their thoughtful insights during our discussions.

\bibliographystyle{ACM-Reference-Format}
\bibliography{references.bib}

\end{document}